\documentstyle[preprint,aps]{revtex} 
\begin{document} 

\draft 

\preprint{UTPT-96-03} 

\title{Linearization Instability of Gravitational Waves Interacting with Matter in
General Relativity}

\author{J. W. Moffat} 

\address{Department of Physics, University of Toronto,
Toronto, Ontario, Canada M5S 1A7} 

\date{\today}

\maketitle 

\begin{abstract}%
The gravitational wave solutions obtained from a perturbation about conformally flat
backgrounds in Einstein gravity are investigated.  A perturbation theory analysis of
the Lesame,
Ellis and Dunsby results, based on a covariant approach, shows that for gravitational
waves interacting with irrotational dust, the equations are linearization
unstable. The gravitational wave equations
based on the Weyl curvature tensor must be solved by
non-perturbative methods. The significance of this result for gravitational wave
calculations and experiments is discussed.
\end{abstract} 

\pacs{ } 

\section{Introduction}

Perturbations of spatially homogeneous and isotropic universes have been
investigated by several authors, beginning with the work of
Lifshitz\cite{Lifshitz}, Bonnor\cite{Bonnor}, Lifshitz and Khalatnikov\cite
{Khalatnikov}, Ehlers\cite{Ehlers}, Hawking\cite{Hawking}, Sachs and
Wolfe\cite{Sachs},
D'Eath\cite{Death} and Moncrief\cite{Moncrief}. More recently, a fully gauge
invariant formulation of perturbation theory has been presented by Bruni, Dunsby,
and Ellis\cite{Ellis1}. By using the lemma of Stewart and Walker\cite{Stewart},
they derived covariant equations from the Bianchi and Ricci identities, which are
gauge invariant with respect to a zero-order conformally flat background, such as
the Friedmann, Robertson, and Walker model (FRW). Following
this work, Lesame, Ellis, and Dunsby\cite{Ellis2,Ellis3} proved that a tetrad
frame exists in which the shear tensor and its covariant time derivative are
diagonalizable, if and only if the divergence of the magnetic Weyl tensor vanishes. 
Moreover, using the Ricci constraints they showed that the magnetic part of the
Weyl tensor also vanishes. This result is applied to a perturbative analysis of the
constraint equations and shows that it leads to a linearization instability in the
perturbative
expansions about a conformally flat spacetime. The
implications for gravitational wave calculations in GR and for future gravitational
wave experiments is considered in the following.

\section{Field Equations and Identities}

We shall assume Einstein's field equations in the form
\begin{equation}
R_{\mu\nu}-\frac{1}{2}g_{\mu\nu}R+\Lambda g_{\mu\nu}=8\pi GT_{\mu\nu},
\end{equation}
where $\Lambda$ is the cosmological constant, and $T_{\mu\nu}$ is the 
stress-energy momentum tensor of the matter, described by a fluid. The Riemann
and Ricci tensors are given by
\begin{equation}
\label{Ricci}
u_{\mu;[\nu\sigma]}=2{R^\alpha}_{\mu\sigma\nu}u_\alpha,\quad R_{\mu\nu}
={{R_\mu}^\alpha}_{\nu\alpha}.
\end{equation}
We have for a perfect fluid:
\begin{equation}
T_{\mu\nu}=\rho u_\mu u_\nu+ph_{\mu\nu},
\end{equation}
where $\rho$ is the density, $p$ is the pressure, $u_\mu$ is the velocity of the fluid,
$u_\mu u^\mu=-1$, and $h_{\mu\nu}=g_{\mu\nu}+u_\mu u_\nu$ is the projection
operator into the hyperplane orthogonal to $u_\mu: h_{\mu\nu}u^\nu=0$.
The gradient of the velocity vector is decomposed as 
\begin{equation}
u_{\mu;\nu}=\omega_{\mu\nu}+\sigma_{\mu\nu}+\frac{1}{3}h_{\mu\nu}\theta
-{\dot u_\mu}u_\nu,
\end{equation}
where ${\dot u_\mu}=u_{\mu;\nu}u^\nu$ is the acceleration,
$\theta={u_{\mu}}^{;\mu}$ is the expansion, $\sigma_{\mu\nu}=u_{(\alpha;\beta)}
h^\alpha_\mu h^\beta_\nu-\frac{1}{3}h_{\mu\nu}\theta$ is the shear, and
$\omega_{\mu\nu}=u_{[\alpha;\beta]}h^\alpha_\mu h^\beta_\nu$ is the rotation of
the flow lines $u_\mu$. 

The Weyl curvature tensor can be decomposed into the Ricci tensor $R_{\mu\nu}$
and the Riemann tensor $R_{\mu\nu\alpha\beta}$:
\begin{equation}
C_{\mu\nu\alpha\beta}=R_{\mu\nu\alpha\beta}+g_{\mu[\beta}R_{\alpha]\nu}
+g_{\nu[\alpha}R_{\beta]\mu}+\frac{1}{3}Rg_{\mu[\alpha}g_{\beta]\nu},
\end{equation}
and $C_{\mu\nu\alpha\beta}=C_{[\mu\nu][\alpha\beta]}$.

The ``electric" and ``magnetic" components of the Weyl tensor are defined by
\begin{equation}
E_{\mu\nu}\equiv E_{(\mu\nu)}=C_{\mu\alpha\nu\beta}u^\alpha u^\beta,
\end{equation}
and
\begin{equation}
H_{\mu\nu}\equiv
H_{(\mu\nu)}=\frac{1}{2}{\epsilon_{\mu\gamma}}^{\alpha\beta}
C_{\alpha\beta\nu\sigma}u^\gamma u^\sigma,
\end{equation}
where $\epsilon_{\mu\nu\alpha\beta}$ is the Levi-Cevita symbol, 
$E_{\mu\nu}$ and $H_{\mu\nu}$ satisfy: $E_{\mu\nu}u^\nu=0,
{E_\mu}^\mu=0$ and $H_{\mu\nu}u^\nu=0, {H_\mu}^\mu=0$. 

The Bianchi identities read:
\begin{equation}
R_{\mu\nu[\alpha\beta;\sigma]}=0,
\end{equation}
while the Ricci identities are given in (\ref{Ricci}).

\section{Propagation and Constraint Equations}

We shall assume that the vorticity, $\omega_{\mu\nu}$, and the pressure,
$p$, are negligible. Then, the Bianchi identities
give Maxwell-type equations\cite{Hawking,Ellis1}:
\begin{mathletters}
\begin{eqnarray}
\label{Eequation}
{h^\nu}_\mu
E_{\nu\alpha;\beta}h^{\alpha\beta}-\epsilon_{\mu\nu\alpha\beta}u^\nu
{\sigma^\alpha}_\rho H^{\beta\rho}&=&\frac{8\pi}{3}G{h^\nu}_\mu\rho_{;\nu},\\
\label{Hequation}
{h^\nu}_\mu H_{\nu\alpha;\beta}h^{\alpha\beta}
-\epsilon_{\mu\nu\alpha\beta}u^\nu{\sigma^\alpha}_\rho
E^{\beta\rho}&=&0,\\
\label{dotE}
{\dot E}_{\mu\nu}+E_{\mu\nu}\theta+
{h^\gamma}_{(\mu}\epsilon_{\nu)\alpha\beta\rho}u^\alpha
{H_\gamma}^{\beta;\rho}-{E^\alpha}_{(\mu}\sigma_{\nu)\alpha}\nonumber \\
-\epsilon_{\mu\alpha\beta\sigma}\epsilon_{\nu\rho\tau\delta}u^\alpha u^\rho
\sigma^{\beta\tau}E^{\sigma\delta}
&=&-4\pi G\rho\sigma_{\mu\nu},\\
\label{dotH}
{\dot H}_{\mu\nu}+H_{\mu\nu}\theta
-{h^\gamma}_{(\mu}\epsilon_{\nu)\alpha\beta\rho}u^\alpha
{E_\gamma}^{\beta;\rho}-{H^\alpha}_{(\mu}\sigma_{\nu)\alpha}\nonumber \\
-\epsilon_{\mu\alpha\beta\sigma}\epsilon_{\nu\rho\tau\delta}u^\alpha u^\rho
\sigma^{\beta\tau}H^{\sigma\delta}&=&0,
\end{eqnarray}
\end{mathletters}
where we have used ${\dot u}_\mu=0$ for $p=0$.

The expansion parameter $\theta$ obeys the Raychaudhuri equation:
\begin{equation}
{\dot \theta}+\frac{1}{3}\theta^2+2\sigma^2+4\pi G\rho=0,
\end{equation}
where $\sigma^2=\frac{1}{2}\sigma^{\mu\nu}\sigma_{\mu\nu}$. The shear
tensor satisfies the evolution equation:
\begin{equation}
{\dot \sigma}_{\mu\nu}+\sigma_{\mu\alpha}{\sigma^\alpha}_\nu
-\frac{2}{3}\theta\sigma_{\mu\nu}+E_{\mu\nu}=0.
\end{equation}

The Ricci constraint equations obtained from (\ref{Ricci}) are given by
\begin{equation}
\label{thetaconstraint}
{h^\mu}_\nu(\frac{2}{3}\theta^{;\nu}-{h^\beta}_\alpha
{\sigma^{\nu\alpha}}_{;\beta})=0,
\end{equation}
and
\begin{equation}
\label{Hconstraint}
H_{\mu\nu}=-{h^\sigma}_\mu {h^\rho}_\nu{\sigma_{(\sigma}}^{\tau;\delta}
\epsilon_{\rho)\kappa\tau\delta}u^\kappa.
\end{equation}

\section{The Background Spacetime and the Constraint Equations}

Since the unperturbed background spacetime is conformally flat, the Weyl tensor
vanishes, as do the projected fields $E_{\mu\nu}$, $H_{\mu\nu}$,
the shear tensor $\sigma_{\mu\nu}$ and the vorticity tensor $\omega_{\mu\nu}$.
Moreover, $u_\mu=\tau_{;\mu}$, where $\tau$ measures the proper time
along the world  lines. The spacetime is homogeneous and isotropic with 
3-surfaces of constant curvature. For the FRW universe, the metric is
\begin{equation}
ds^2=-d\tau^2+R^2d\sigma^2,
\end{equation}
where $R=R(\tau)$ and $d\sigma^2$ is the line element of space with zero,
unit positive or negative curvature.

It has been shown by Hawking\cite{Hawking} that for $p=0$ the rotation satisfies:
$\omega=\omega_0/R^2$, so that
the rotation dies away as the universe expands, in accord with our assumption that
we can neglect
$\omega_{\mu\nu}$.

Let us define
\[
\Sigma_{\mu\beta\rho}=\epsilon_{\mu\nu\alpha\beta}u^\nu{\sigma_\rho}^\alpha.
\]
We can write equation (\ref{Eequation}) and (\ref{Hequation}) as
\begin{equation}
(\hbox{div}\,E)_\mu-\Sigma_{\mu\beta\sigma}H^{\beta\sigma}=
\frac{8\pi}{3}G{h^\nu}_\mu\rho_{;\nu},
\end{equation}
and
\begin{equation}
(\hbox{div}\,H)_\mu-\Sigma_{\mu\beta\sigma}E^{\beta\sigma}=0.
\end{equation}
By considering the proper time derivative of the constraint equation
(\ref{thetaconstraint}):
\begin{equation}
h^{\mu\nu}[\frac{2}{3}(\theta_{;\nu})^{.}
-{h^\beta}_\alpha({\sigma^\alpha}_{\nu;\beta})^{.}]=0,
\end{equation}
and using a tetrad frame, Lesame et al.,\cite{Ellis2} obtained:
\begin{equation}
\sigma_1H_1=\sigma_2H_2=\sigma_3H_3,\quad (1+\lambda+\lambda^2)\sigma
H=0,
\end{equation}
where we used the notation $\sigma_i=\sigma_{ii}$ (no sum), $H_i=H_{ii}$ 
(no sum) ($i=1,2,3$) and $\lambda$ is a constant. Moreover, 
for the diagonal case:
\[
\sigma_1=\sigma, \quad \sigma_2=\lambda\sigma,
\]
and
\[
H_1=\lambda H,\quad H_2=H.
\]
Also, the trace-free condition: ${H_\mu}^\mu=0$ gives
\[
\sigma_3=-(1+\lambda)\sigma,\quad H_3=-(1+\lambda)H.
\]

For arbitrary $\lambda$ and $\sigma\not= 0$, it follows that $H=0$. For
$1+\lambda+\lambda^2=0$, the values of $\lambda$ are complex, but in tetrad
form:
\[
\vert\sigma\vert^2=\frac{1}{2}(1+\lambda+\lambda^2)\sigma^2=0,\quad 
\vert H\vert^2=(1+\lambda+\lambda^2)H^2=0
\]
so both the shear and the magnetic Weyl tensor are zero, which also follows for
the case when $\sigma=0$, since this implies that the spacetime is FRW for
which $E=H=0$. This leads to the Lesame, Ellis and Dunsby 
theorem\cite{Ellis2,Ellis3}: 
{\obeylines\smallskip
For irrotational dust, the divergence of the magnetic Weyl tensor vanishes:
$(\hbox{div}\,H)_\mu=0$ for a shear tensor that is diagonalizable in a tetrad
frame, if and only if the magnetic Weyl tensor vanishes, $H_{\mu\nu}=0$.
\smallskip}

\section{Linear Gravitational Wave Equation}

We shall now expand the fields $E_{\mu\nu},
H_{\mu\nu}$ and the tensor $\sigma_{\mu\nu}$ in a power series
with respect to a small parameter $\epsilon << 1$:
\begin{mathletters}
\begin{eqnarray}
E_{\mu\nu}&=&\epsilon E^{(1)}_{\mu\nu}+
\epsilon^2 E^{(2)}_{\mu\nu}+O(\epsilon^3),\\
H_{\mu\nu}&=&\epsilon H^{(1)}_{\mu\nu}+
\epsilon^2 H^{(2)}_{\mu\nu}+O(\epsilon^3),\\
\sigma_{\mu\nu}&=&\epsilon\sigma_{\mu\nu}^{(1)}+\epsilon^2
\sigma_{\mu\nu}^{(2)}+O(\epsilon^3).
\end{eqnarray}
\end{mathletters}

Let us consider the first-order equations resulting from Eqs. (\ref{dotE}) and
(\ref{dotH}):
\begin{eqnarray}
\label{lindotE}
{\dot E}_{\mu\nu}^{(1)}+E_{\mu\nu}^{(1)}\theta+
{h^\gamma}_{(\mu}\epsilon_{\nu)\alpha\beta\rho}u^\alpha
{H_\gamma}^{(1)\beta;\rho}
&=&-4\pi G\rho\sigma_{\mu\nu}^{(1)},\\
\label{lindotH}
{\dot H}_{\mu\nu}^{(1)}+H^{(1)}_{\mu\nu}\theta
-{h^\gamma}_{(\mu}\epsilon_{\nu)\alpha\beta\rho}u^\alpha
{E_\gamma}^{(1)\beta;\rho}&=&0.
\end{eqnarray}
By multiplying (\ref{lindotE}) by $\partial/\partial\tau=u^\alpha\nabla_\alpha$ and
(\ref{lindotH}) by ${h^\mu}_{(\rho}{\epsilon_\beta)}^{\gamma\nu\tau}u_\gamma
\nabla_\tau$, we get\cite{Hawking}
\begin{equation}
\label{waveequation}
{\ddot E}_{\mu\nu}^{(1)}-\Delta^2E^{(1)}_{\mu\nu}
+\frac{7}{3}{\dot E}_{\mu\nu}^{(1)}\theta
+E^{(1)}_{\mu\nu}({\dot\theta}+\frac{4}{3}\theta^2
+\frac{8\pi}{3}G\rho)+8\pi G\sigma^{(1)}_{\mu\nu}(\frac{1}{3}\rho
+\frac{1}{2}{\dot\rho})=0,
\end{equation}
where 
\[
\Delta^2E_{\mu\nu}=(E^{(1)}_{\alpha\beta;\tau}{h^\alpha}_\delta{h^\beta}_\rho
{h^\tau}_\gamma)_{;\sigma}h^{\gamma\sigma}{h^\delta}_\mu {h^\rho}_\nu
\]
is the Laplacian operator in the hypersurface $\tau=\hbox{constant}$. It has the
eigenfunction expansion:
\[
E_{\mu\nu}=\Sigma_i a^{(i)}K_{\mu\nu}^{(i)}
\]
with ${\dot K}^{(i)}_{\mu\nu}=0$.

For a non-expanding congruence $u^\alpha$ and empty Minkowski spacetime, Eq.
(\ref{waveequation}) reduces to the wave equation
\begin{equation}
\label{standardwave}
\Box E_{\mu\nu}=0.
\end{equation}

\section{Linearization Instability}

Equating coefficients of powers of $\epsilon$, we obtain from Eqs.
(\ref{Eequation}) and (\ref{Hequation}):
\begin{mathletters}
\begin{eqnarray}
(\hbox{div}\,E^{(1)})_\mu&=&\frac{8\pi}{3}G{h^\nu}_\mu\rho_{;\nu},\\
\label{divHequation}
(\hbox{div}\,H^{(1)})_\mu&=&0,\\
\label{2order}
(\hbox{div}\,E^{(2)})_\mu-\Sigma^{(1)}_{\mu\nu\sigma}H^{(1)\nu\sigma}
&=&\frac{8\pi}{3}G{h^\nu}_\mu\rho_{;\nu},\\
(\hbox{div}\,H^{(2)})_\mu-\Sigma^{(1)}_{\mu\nu\sigma}
E^{(1)\nu\sigma}&=&0,\\
&\vdots& \nonumber \\
(\hbox{div}\,E^{(n)})_\mu-\sum_{i=1}^{n-1}\Sigma^{(i)}_{\mu\nu\sigma}
H^{(n-i)\nu\sigma}&=&\frac{8\pi}{3}G{h^\nu}_\mu\rho_{;\nu},\\
(\hbox{div
}\,H^{(n)})_\mu-\sum_{i=1}^{n-1}\Sigma^{(i)}_{\mu\nu\sigma}
E^{(n-i)\nu\sigma}&=&0,
\end{eqnarray}
\end{mathletters}
and from Eqs. (\ref{dotE}) and (\ref{dotH}):
\begin{mathletters}
\begin{eqnarray}
{\dot E}^{(1)}_{\mu\nu}+E^{(1)}_{\mu\nu}\theta+
{h^\gamma}_{(\mu}\epsilon_{\nu)\alpha\beta\rho}u^\alpha
{H_\gamma}^{(1)\beta;\rho}
&=&-4\pi G\rho\sigma^{(1)}_{\mu\nu},\\
{\dot H}^{(1)}_{\mu\nu}+H^{(1)}_{\mu\nu}\theta
-{h^\gamma}_{(\mu}\epsilon_{\nu)\alpha\beta\rho}u^\alpha
{E_\gamma}^{(1)\beta;\rho}&=&0,\\
{\dot E}^{(2)}_{\mu\nu}+E^{(2)}_{\mu\nu}\theta+
{h^\gamma}_{(\mu}\epsilon_{\nu)\alpha\beta\rho}u^\alpha
{H_\gamma}^{(2)\beta;\rho}-{E^{(1)\alpha}}_{(\mu}{\sigma^{(1)}}_{\nu)\alpha}\nonumber \\
-\epsilon_{\mu\alpha\beta\sigma}\epsilon_{\nu\rho\tau\delta}u^\alpha u^\rho
\sigma^{(1)\beta\tau}E^{(1)\sigma\delta}
&=&-4\pi G\rho\sigma^{(2)}_{\mu\nu},\\
{\dot H}^{(2)}_{\mu\nu}+H^{(2)}_{\mu\nu}\theta
-{h^\gamma}_{(\mu}\epsilon_{\nu)\alpha\beta\rho}u^\alpha
{E_\gamma}^{(2)\beta;\rho}-{H^{(1)\alpha}}_{(\mu}{\sigma^{(1)}}_{\nu)\alpha}\nonumber \\
-\epsilon_{\mu\alpha\beta\sigma}\epsilon_{\nu\rho\tau\delta}u^\alpha u^\rho
\sigma^{(1)\beta\tau}H^{(1)\sigma\delta}&=&0,\\
&\vdots& \nonumber \\
{\dot E}^{(n)}_{\mu\nu}+E^{(n)}_{\mu\nu}\theta+
{h^\gamma}_{(\mu}\epsilon_{\nu)\alpha\beta\rho}u^\alpha
{H_\gamma}^{(n)\beta;\rho}-\sum^{n-1}_{i=1}{E^{(i)\alpha}}_{(\mu}{\sigma^{(n-i)}}_{\nu)\alpha}
\nonumber \\
-\epsilon_{\mu\alpha\beta\sigma}\epsilon_{\nu\rho\tau\delta}u^\alpha u^\rho
\sum_{i=1}^{n-i}\sigma^{(i)\beta\tau}E^{(n-i)\sigma\delta}
&=&-4\pi G\rho\sigma^{(n)}_{\mu\nu},\\
{\dot H}^{(n)}_{\mu\nu}+H^{(n)}_{\mu\nu}\theta
-{h^\gamma}_{(\mu}\epsilon_{\nu)\alpha\beta\rho}u^\alpha
{E_\gamma}^{(n)\beta;\rho}
-\sum_{i=1}^{n-1}{H^{(i)\alpha}}_{(\mu}{\sigma^{(n-i)}}_{\nu)\alpha}
\nonumber \\
-\epsilon_{\mu\alpha\beta\sigma}\epsilon_{\nu\rho\tau\delta}u^\alpha u^\rho
\sum_{i=1}^{n-1}\sigma^{(i)\beta\tau}H^{(n-i)\sigma\delta}&=&0.
\end{eqnarray}
\end{mathletters}

By inserting the infinite power series in $\epsilon$ for $H_{\mu\nu}$ into the
result of the Lesame, Ellis and Dunsby theorem: $(\hbox{div}H)_\mu=0$
and $H_{\mu\nu}=0$, we find that $H_{\mu\nu}$ vanishes in every order.
Then, Eq. (\ref{divHequation}) leads to the result: 
\begin{equation}
H_{\mu\nu}^{(1)}= 0,
\end{equation}
and to first-order we get
\begin{mathletters}
\begin{eqnarray}
(\hbox{div}\,E^{(1)})_\mu
&=&\frac{8\pi}{3}G{h^\nu}_\mu\rho_{;\nu},\\
{\dot E}^{(1)}_{\mu\nu}+E^{(1)}_{\mu\nu}\theta&=&-4\pi
G\rho\sigma^{(1)}_{\mu\nu},\\
\hbox{curl}\,E^{(1)}_{\mu\nu}&=&{h^\gamma}_{(\mu}\epsilon_{\nu)\alpha
\beta\rho}
u^\alpha{E_\gamma}^{(1)\beta;\rho}=0.
\end{eqnarray}
\end{mathletters}
In second-order, we obtain
\begin{mathletters}
\begin{eqnarray}
(\hbox{div}\,E^{(2)})_\mu
&=&\frac{8\pi}{3}G{h^\nu}_\mu\rho_{;\nu},\\
(\hbox{div}\,H^{(2)})_\mu-\Sigma^{(1)}_{\mu\nu\sigma}
E^{(1)\nu\sigma}&=&0,\\
{\dot E}^{(2)}_{\mu\nu}+E^{(2)}_{\mu\nu}\theta+
{h^\gamma}_{(\mu}\epsilon_{\nu)\alpha\beta\rho}u^\alpha
{H_\gamma}^{(2)\beta;\rho}-E^{(1){\alpha}}_{(\mu}{\sigma^{(1)}}_{\nu)\alpha}\nonumber \\
-\epsilon_{\mu\alpha\beta\sigma}\epsilon_{\nu\rho\tau\delta}u^\alpha u^\rho
\sigma^{(1)\beta\tau}E^{(1)\sigma\delta}
&=&-4\pi G\rho\sigma^{(2)}_{\mu\nu},\\
{\dot H}^{(2)}_{\mu\nu}+H^{(2)}_{\mu\nu}\theta
-{h^\gamma}_{(\mu}\epsilon_{\nu)\alpha\beta\rho}u^\alpha
{E_\gamma}^{(2)\beta;\rho}&=&0.
\end{eqnarray}
\end{mathletters}%
Thus, neither the first-order nor the second-order of perturbation theory predict the
same evolution
as the rigorous theory. This means that there is a linearization instability and the
gravitational wave equations can only be solved by non-perturbative methods.  It is
not possible to derive the standard linear wave equation (\ref{standardwave})
in empty spacetime, because of the absence of a $\hbox{curl}\,H^{(1)}$ term in the
first-order equations. This confirms the Lesame et al., conclusions\cite{Ellis2}.
However, our analysis reveals that the breakdown of the linearization scheme
follows inevitably from the Lesame, Ellis and Dunsby theorem, when a power series
solution of the equations is sought.

\section{Consequences for Gravitational Wave Experiments}

Considerable work has been devoted to studying the linearization stability of
gravitational wave perturbation theory in Minkowski 
spacetime\cite{Moncrief,Fischer,Choquet}. It
demonstrated that Cauchy stability can be proved and that the linear approximation
of the field equations can be expected to be stable for the empty space Einstein
equations. However, the problem of perturbative stability for cosmological
solutions, such as the FRW solution of the Einstein field equations, is
a more difficult issue to resolve. The full Einstein field equations have (under
certain conditions)
a Cauchy solution which is stable, but potential difficulties arise when
a perturbative analysis of the equations is performed due to the role of the
Bianchi and Ricci constraints. This is borne out by the results of Lesame et al.,
who consistently retain the Ricci constraints in their analysis by demanding that
they hold in the time evolution of the initial data.
A central issue is that the Lesame et al., analysis is based
on a fully gauge invariant (covariant) perturbation theory. As first stressed by 
Hawking\cite{Hawking}, a non-gauge invariant analysis based on metric
perturbations is not reliable and can yield misleading results. He therefore
suggested that the perturbation theory be based on an analysis of the curvature
tensors.

In the proposed gravitational wave experiments such as the LIGO/VIRGO or
LISA projects\cite{Thorne}, which attempt to detect gravitational radiation directly, 
the gravitational waves arrive at the detector after travelling through the universe
described by a cosmological model such as the standard FRW model. In these
models there is always a non-vanishing density $\rho$ i.e., the estimated density
of the smoothed out fluid of the universe is
$\rho\sim 10^{-29}\,\hbox{g}\,\hbox{cm}^{-3}$. Even though this is
a small density it will play an important role in the perturbative stability analysis,
because the linearization instability of the gravitational wave equations in the
presence
of an irrotational dust is expected to be singular, in the sense that the solutions of the
gravitational wave equations become singular in the limit of a conformally
flat FRW solution. Thus, only non-pertubative or non-linear numerical solutions
of the gravitational wave equations, based on a gauge invariant set of equations
within a cosmological model scenario, can be trusted to produce reliable physical
predictions for gravitational wave experiments. 

The proposed strong
gravitational wave sources which one hopes can produce detectable radiation,  
such as the coalescence of neutron star and stellar mass black hole binaries in
distant galaxies or colliding black holes, involve strong gravitational
fields with non-vanishing matter and shear in the close vicinity of the sources,
requiring a non-perturbative solution of the wave equations.

Although the conclusions drawn about the linearization instability of the
cosmological gravitational wave equations by Lesame et al., and in the present
work, is based
on irrotational dust, such an approximation is expected to be very reliable in the
present universe, since the vorticity and pressure are vanishingly small.
Therefore, the linearization instability of the gravitational wave solutions travelling
through a medium discussed here, {\it plays an important role for realistic
gravitational wave calculations and experiments in the present universe.}

Perhaps, this kind of linearization instability of the gravitational wave equations,
arises from a breaking of the conformal symmetry of the
cosmological model e.g., the FRW model which is conformally invariant
with a constant curvature. Further work should be carried out to understand the
mechanism associated with this symmetry breaking, which is responsible
for producing the linearization instability. This mechanism may be related to
the result obtained by Kundt and Tr\"umper and Szekeres\cite{Kundt,Szekeres},
that in an exact treatment of GR, no Petrov type N irrotational dust solutions exist.

\section{Conclusions}

A perturbative analysis of the Weyl curvature Maxwell-type equations and the
associated constraints about a
conformally flat background, which was chosen to be represented by the FRW
model, led to a linearization instability when the Lesame, Ellis and Dunsby
theorem was invoked. This confirmed the conclusions about linearization instability
inferred by Lesame et al.,\cite{Ellis2,Ellis3}.

It is important to emphasize that this instability is {\it inevitable}
already at first-order in a conventional power series expansion of the Weyl
curvature equations in a small parameter $\epsilon$, and reflects itself in the
higher order equations. The Ricci constraints must be imposed at the lowest order,
which results in $H^{(1)}_{\mu\nu}$ vanishing and the breakdown of the power
series perturbation theory. Thus, the perturbative instability is a {\it generic feature}
of any consistent expansion about an FRW backround, based on the covariant
Weyl curvature approach. Perhaps, a reinvestigation of the Cauchy perturbative
instability of gravitational wave equations, expanded about an irrotational dust
solution, is needed to fully understand the implications of this result.

Since gravitational wave experiments are conducted for strong gravitational field
sources with non-vanishing matter density and shear, and since the gravitational
waves that the experiments hope to detect travel through an irrotational dust in the
present universe, then the whole issue of the validity of the perturbative calculations
of gravitational wave solutions should be treated with caution. This may
become an important topic of study in numerical calculations of gravitational waves
in the near-linear regime\cite{Seidel}.

\acknowledgments

I thank G. Ellis, M. Bruni, M. Clayton, J. L\'egar\'e and P. Savaria for helpful and
stimulating
discussions. This work was supported by the Natural Sciences and Engineering
Research
Council of Canada.

\end{document}